\documentclass[preprint,aps,pre,amsmath,amssymb,superscriptaddress]{revtex4-1}
\usepackage[T1]{fontenc}
\usepackage[utf8]{inputenc}
\usepackage{newtxtext}
\usepackage[subscriptcorrection,varg]{newtxmath}
\usepackage[cal=cm]{mathalfa}
\usepackage[dvipsnames]{xcolor}
\usepackage[colorlinks=true,allcolors=Blue,breaklinks]{hyperref}
\usepackage{siunitx,graphicx,physics}
\usepackage[capitalize,noabbrev]{cleveref}

\makeatletter
\def\fps@figure{htb!}
\makeatother

\begin{document}

\title{Edge Shear Flows and Particle Transport near the Density Limit of the HL-2A Tokamak}
\date{\today}

\affiliation{Center for Energy Research, University of California San Diego, La Jolla, CA 92093, USA}
\affiliation{Center for Astrophysics and Space Sciences, University of California San Diego, La Jolla, CA 92093, USA}
\affiliation{Center for Fusion Sciences, Southwestern Institute of Physics, Chengdu, Sichuan 610041, China}
\affiliation{Department of Engineering Physics, Tsinghua University, Beijing 100084, China}
\affiliation{School of Physical Sciences, University of Science and Technology of China, Hefei, Anhui 230026, China }

\author{R.~Hong}
\affiliation{Center for Energy Research, University of California San Diego, La Jolla, CA 92093, USA}

\author{G.~R.~Tynan}
\affiliation{Center for Energy Research, University of California San Diego, La Jolla, CA 92093, USA}
\affiliation{Center for Fusion Sciences, Southwestern Institute of Physics, Chengdu, Sichuan 610041, China}

\author{P.~H.~Diamond}
\affiliation{Center for Astrophysics and Space Sciences, University of California San Diego, La Jolla, CA 92093, USA}
\affiliation{Center for Fusion Sciences, Southwestern Institute of Physics, Chengdu, Sichuan 610041, China}

\author{L.~Nie}
\author{D.~Guo}
\author{T.~Long}
\affiliation{Center for Fusion Sciences, Southwestern Institute of Physics, Chengdu, Sichuan 610041, China}

\author{R.~Ke}
\affiliation{Center for Fusion Sciences, Southwestern Institute of Physics, Chengdu, Sichuan 610041, China}
\affiliation{Department of Engineering Physics, Tsinghua University, Beijing 100084, China}

\author{Y.~Wu}
\author{B.~Yuan}
\affiliation{Center for Fusion Sciences, Southwestern Institute of Physics, Chengdu, Sichuan 610041, China}
\affiliation{School of Physical Sciences, University of Science and Technology of China, Hefei, Anhui 230026, China }

\author{M.~Xu}
\author{the HL-2A Team}
\affiliation{Center for Fusion Sciences, Southwestern Institute of Physics, Chengdu, Sichuan 610041, China}

\begin{abstract}
{\normalsize
Edge shear flow and its effect on regulating turbulent transport have long been suspected to play an important role in plasmas operating near the Greenwald density limit $ n_G $.
In this study, equilibrium profiles as well as the turbulent particle flux and Reynolds stress across the separatrix in the HL-2A tokamak are examined as $ n_G $ is approached in ohmic L-mode discharges.
As the normalized line-averaged density $ \bar{n}_e/n_G $ is raised, the shearing rate of the mean poloidal flow $ \omega_{\rm sh} $ drops, and the turbulent drive for the low-frequency zonal flow (the Reynolds power $ \mathcal{P}_{Re} $) collapses.
Correspondingly, the turbulent particle transport increases drastically with increasing collision rates.
The geodesic acoustic modes (GAMs) gain more energy from the ambient turbulence at higher densities, but have smaller shearing rate than low-frequency zonal flows.
The increased density also introduces decreased adiabaticity which not only enhances the particle transport but is also related to reduction in the eddy-tilting and the Reynolds power.
Both effects may lead to cooling of edge plasmas and therefore the onset of MHD instabilities that limit the plasma density.
}
\end{abstract}
\maketitle

\section{Introduction}

High density operations are desirable in magnetic confinement fusion reactors.
Densities of the order of $ 10^{20} $ \si{\per\cubic\meter} are required for achieving ignition  \cite{ITERPEG1999NF2137,Greenwald2002PPaCF27,Zohm2013NF73019}.
However, raising the line-averaged density, $ \bar{n}_e $, to the Greenwald limit, $ n_{G}\, [10^{20}\,\si{\per\cubic\meter}] = \frac{I_{p}[\si{\mega\ampere}]}{\pi a^2 [\si{\square\meter}]}$, usually leads to a significant reduction in confinement time, or even disruption, when $ n_{G} $ is exceeded \cite{Greenwald2002PPaCF27,Greenwald1988NF2199}.
Since the discovery of this density limit, extensive experimental studies have shown that the Greenwald limit can be exceeded by increasing the core density while keeping the edge density low, i.e.~by operating with peaked density profiles, using optimized fueling techniques \cite{Kamada1991NF1827,Lang2012NF23017,Mahdavi2002NF52,Valovic2002PPaCF1911}.
These findings provide strong evidence linking the density limit to the edge physics \cite{Greenwald2002PPaCF27}.

Among the phenomena in the plasma boundary region, edge cooling and radiation loss are found to be associated with the density limit, and have been widely investigated \cite{Greenwald2002PPaCF27,Greenwald1988NF2199,Suttrop1997NF119,Wesson1989NF641}.
In radiation models \cite{Greenwald2002PPaCF27,Connor2002PPaCF121}, the radiative heat loss due to increased impurity content in the plasma is thought to dominate the power balance at high densities, resulting in strong edge cooling and thus an increased resistivity, causing the toroidal current channel to shrink.
The current shrinkage then leads to an increased current density gradient and the onset of resistive MHD instabilities.
In particular, a thermo-resistive tearing mode model \cite{Gates2012PRL165004,Gates2013NF63008,Gates2016PoP56113}, in which radiative cooling is balanced with the ohmic heating inside magnetic islands, has been invoked to explain the dependence of the current density in the Greenwald limit scaling.

Although radiation models have achieved some success in explaining the empirical scaling, they do not address the mechanism that initiates edge cooling.
One likely candidate is enhanced edge transport, i.e.~turbulent particle and heat fluxes \cite{Greenwald2002PPaCF27}.
We also note that enhanced particle transport has been observed in advance of any change in the MHD activity in both experiments and numerical simulations when $n_G$ is approached \cite{Brower1991PRL200,Rogers1998PRL4396,LaBombard2001PoP2107,LaBombard2005NF1658,Suttrop1999JoNM118}.
At fixed pressure, the higher density usually implies reduced temperature and hence increases dissipative and resistive effects.
These effects destabilize the resistive modes, which lead to enhanced fluctuation levels and turbulent transport \cite{Thayer1987PoF3724}.
Also, turbulent transport can be suppressed or mitigated by zonal flows that are in turn driven by the turbulence via the Reynolds force \cite{Diamond2005PPaCF35,Fujisawa2009NF13001,Tynan2009PPaCF113001,Manz2009PRL165004,Xu2012PRL245001,Xu2011PRL55003,Stroth2011PPaCF24006,Birkenmeier2013PRL145004}.
This self-regulating process has been recognized as an important mechanism for the \textit{L--H} transition that leads to the edge transport barrier and improved plasma confinement \cite{Cziegler2015NF83007,Cziegler2014PPaCF75013,Tynan2013NF73053,Yan2014PRL125002,Tynan2016PPaCF44003}.
However, zonal flows are subject to strong collisional damping \cite{Diamond2005PPaCF35}.
Weaker zonal flows cannot efficiently trigger the "tilt-stretch-absorption" process \cite{Manz2009PRL165004,Xu2011PRL55003}, and therefore result in reduced Reynolds force.
As a result, the self-regulation process is inhibited when the density limit is approached, and edge turbulent fluxes should increase.
The competition between collisionality triggered instabilities and the stabilizing effects of $ \mathbf{E\times B} $ shear flows may lead to the limit of pedestal density.

Studies of turbulent transport in the scrape-off layer (SOL) as $ \bar{n}_e $ is raised to the Greenwald limit have shown a pronounced increase in SOL turbulence intermittency \cite{LaBombard2001PoP2107,LaBombard2005NF1658}, demonstrating that turbulent transport undergoes important changes as the density limit is approached.
However, as of now, to our knowledge, the evolution of the turbulent particle and momentum fluxes, zonal flows and GAMs, and their interactions across the SOL, separatrix and edge plasma region have not been reported.

In present study we examine the behavior of the edge shear flows and cross-field particle transport as $ \bar{n}_e $ approaches the Greenwald limit in ohmic HL-2A tokamak plasmas.
The discharges and the diagnostic tools used for this work are discussed in \cref{sec:setup}.
The experimental results and discussions on the evolution of edge shear flows, the nonlinear energy transfer, and the edge particle transport are presented in \cref{sec:result} and \cref{sec:disc}, respectively.
A summary of this work is given in \cref{sec:concl}.

\section{Experimental Arrangement}
\label{sec:setup}

The experiment was carried out in the HL-2A tokamak \cite{Liu2005NF239,Duan2013NF104009,Xu2012PRL245001}, which has a major radius of $R=\SI{1.65}{\meter}$ and a minor radius of $a=\SI{0.4}{\meter}$.
In this study ohmic deuterium plasmas were produced in the lower-single-null (LSN) geometry with the `favorable' $ \mathbf{\nabla B \times B} $ drift direction (toward the X-point).
The plasma current was $ I_p=\SI{150}{\kilo\ampere} $, the toroidal magnetic field was $ B_T=\SI{1.3}{\tesla}$, and the edge safety factor was about $ 3.5-4 $.
The Greenwald limit density was $ n_G = I_p/\pi a^2 \approx 3.2 \times 10^{19}$ \si{\per\cubic\meter} in these conditions.
In this shot-by-shot density scanning experiment, the line-averaged densities $ \bar{n}_e $ measured by the HCN laser interferometer ramped from \numrange{0.8e19}{2.8e19} \si{\per\cubic\meter} which correspond to a normalized density range of $ 0.25 - 0.9 \, n_G $.

A multi-tip Langmuir probe array was used to investigate the edge turbulence and shear flows at the low-field-side (LFS) mid-plane of the tokamak \cite{Xu2012PRL245001}.
The probe is composed of a $ 3 \times 5 $ array of graphite tips, i.e. 5 steps with 3 tips on each step.
The distance between two adjacent tips is 5 mm in the poloidal direction and 2.5 mm in the radial direction.
Tips on the first, the third, and the fifth step were operated as triple probes, providing the electron density $ n_e $ and temperature $ T_e $, as well as the plasma potential $ \phi_p = \phi_f + 2.8 T_e $.
Other tips were used to measure the floating potentials $ \phi_f $.
All probe data were sampled at 1 MHz using 12-bit digitizers.
With this probe setup, we are also able to simultaneously measure the Reynolds stress, $ -\langle \tilde{E}_r \tilde{E}_\theta \rangle / B^2_{T} $, and the turbulent particle flux, $ \Gamma_r = \langle \tilde{n}_e \tilde{E}_{\theta} \rangle/B_{T} $, where $ \tilde{E} = - \nabla \tilde{\phi}_f $.
In previous experiments, the broadband turbulence was found to have a frequency range of $ 30<f<80 $ kHz.
In this study, a fifth-order bandpass Butterworth filter was used to obtain the high-frequency fluctuation signals (20-100 kHz).

\section{Results}
\label{sec:result}

\subsection{Equilibrium Profiles}

\begin{figure}
\includegraphics[width=3in]{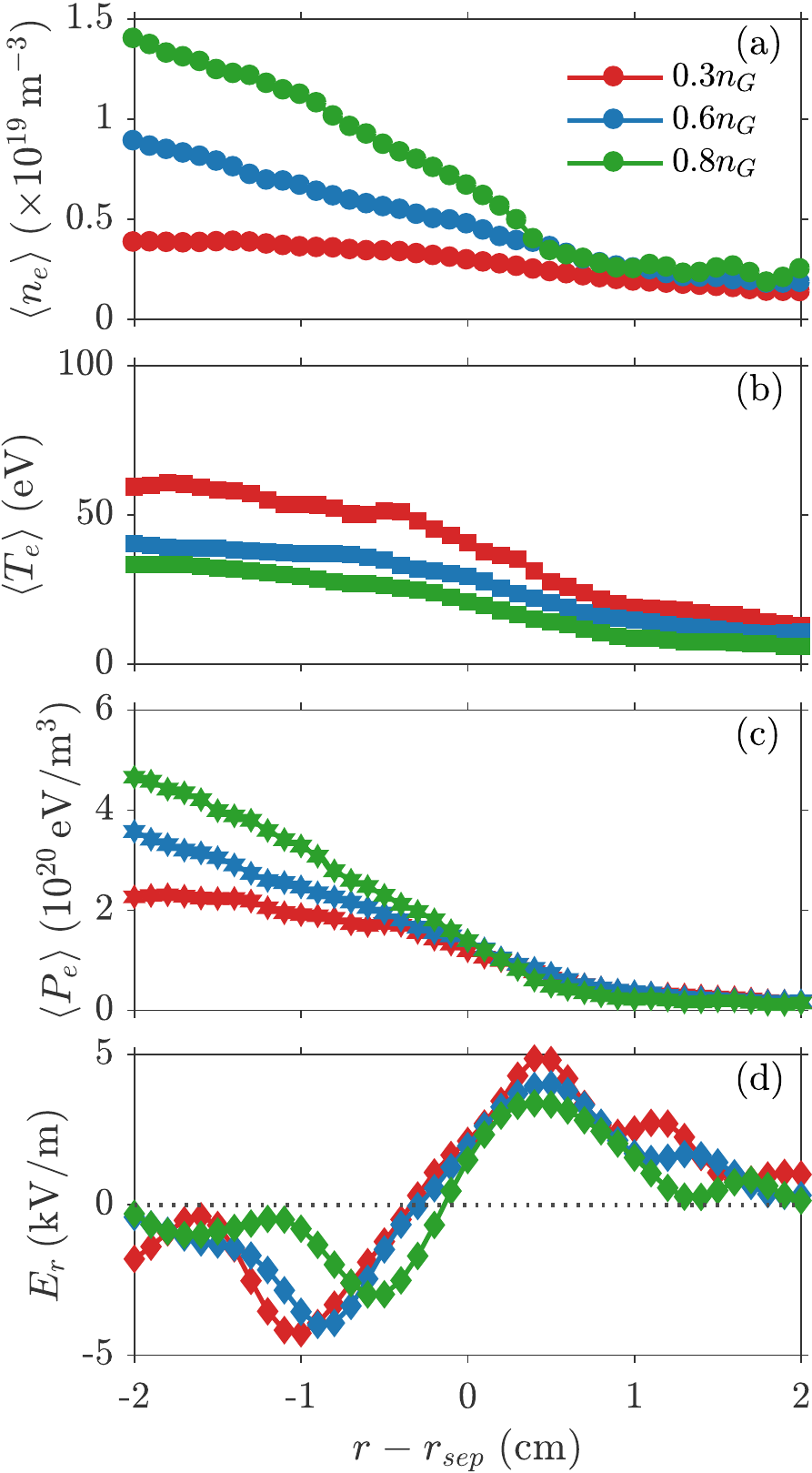}
\caption{\label{fig:profile} Equilibrium profiles of the electron density (a), electron temperature (b), electron pressure (c) and radial electric field (d), at three normalized line-averaged densities $ \bar{n}_e $.}
\end{figure}

Figure \ref{fig:profile} shows the equilibrium profiles of the electron density $ n_e $, electron temperature $ T_e $, electron pressure $ P_e=n_e T_e $, and radial electric field $ E_r = -\partial_r \phi_p $ at three different normalized densities, i.e. $ \bar{n}_e / n_G \approx $ 0.3, 0.6 and 0.8.
These profiles are obtained by taking the time average with 2 millisecond windows.
As the normalized core density, $ \bar{n}_e / n_G $, is raised from 0.3 to 0.8, the edge electron density increases by a factor of 3 at a position about $ 2\,\mathrm{cm} $ inside the separatrix, while the electron temperature drops from about 60 eV to 30 eV.
The electron pressure and its radial gradient increase with $ \bar{n}_e / n_G $.
The peak value of the radial electric field is reduced (\cref{fig:profile}(d)) due to the flattening of the plasma potential profiles at higher $ \bar{n}_e / n_G $ values.
The position of the separatrix is obtained from the magnetic equilibrium reconstruction.

\subsection{Kinetic Energy Transfer Analysis}
\label{subsec:eng}

\begin{figure}
\includegraphics[width=3in]{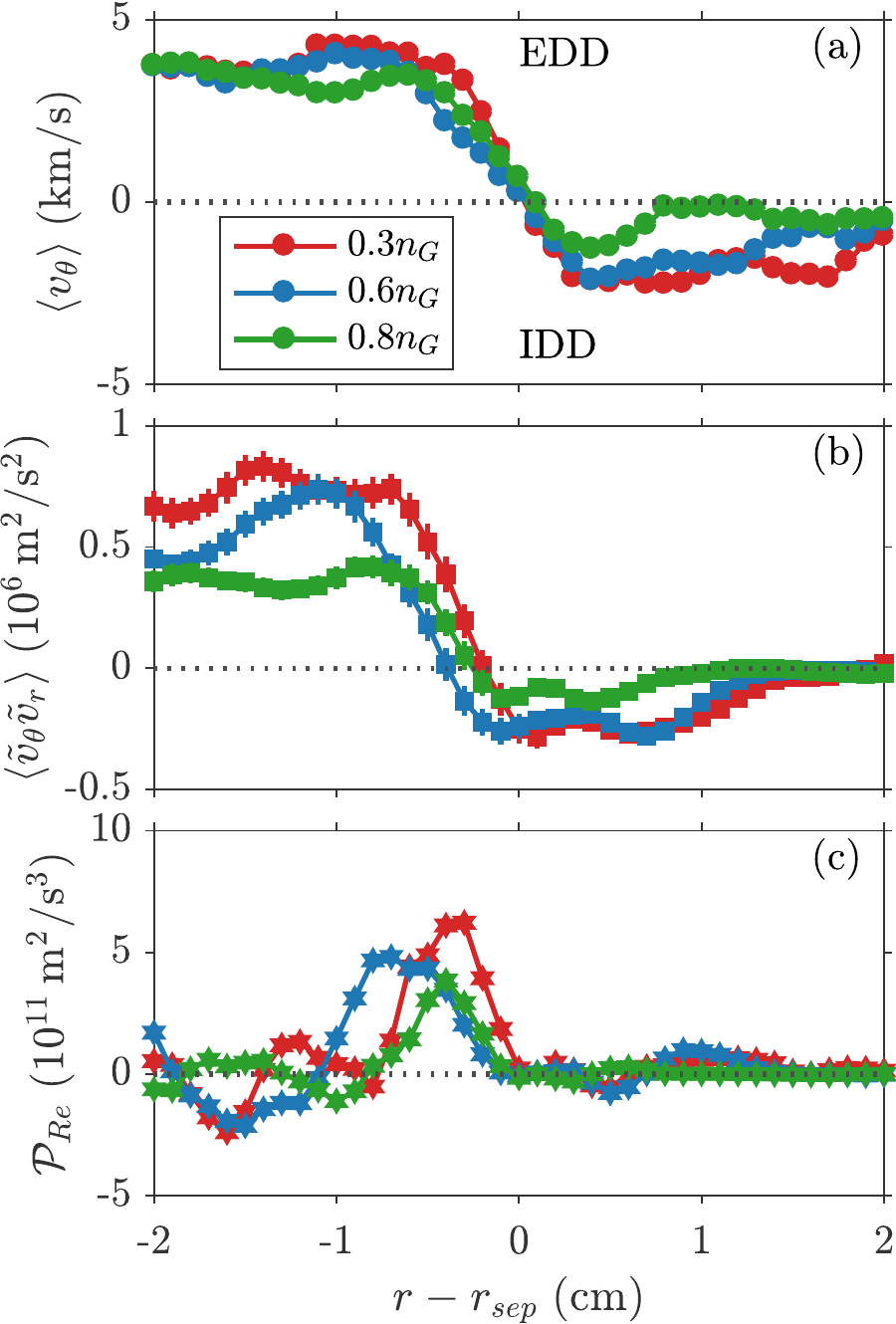}
\caption{\label{fig:vphase} Profiles of the poloidal phase velocity $ \langle v_\theta \rangle $ (a), turbulent Reynolds stress $ \langle \tilde{v}_\theta \tilde{v}_r \rangle $ (b) and Reynolds power $ \mathcal{P}_{Re} = - \langle v_\theta \rangle \partial_r \langle \tilde{v}_\theta \tilde{v}_r \rangle $ (c), at three normalized line-averaged densities $ \bar{n}_e / n_G $. The positive $ \langle v_\theta \rangle $ is in electron diamagnetic drift (EDD) direction and negative $ \langle v_\theta \rangle $ in ion diamagnetic drift (IDD) direction.}
\end{figure}

The poloidal phase velocity of plasma fluctuations, $ \langle v_\theta \rangle $ (\cref{fig:vphase}(a)), can be inferred using the time-delay estimation (TDE) technique from two poloidally separated floating potential signals \cite{Xu2012PRL245001}.
Here, a pair of 2 msec long time series are used to evaluate the local dispersion relations, i.e.~conditional power spectra $ S(k_\theta|f) = \frac{S(k_\theta,f)}{\sum_{k_{\theta}} S(k_{\theta},f)} $ at each position, introducing a spatial resolution of 1 mm (with a 50\% overlap).
This corresponds to the distance over which the probe tips move during the 2 msec window.
The high-frequency fluctuations appear to propagate in the electron diamagnetic drift (EDD) direction inside the separatrix and propagate in the ion diamagnetic drift (IDD) direction in the SOL region.
When $ \bar{n}_e / n_G $ is raised, $ \langle v_\theta \rangle $ decreases, particularly in the SOL region.
In addition, as shown in \cref{fig:vphase}(b), the turbulent Reynolds stress (with $ \tilde{v}_\theta $ and $ \tilde{v}_r $ in the frequency range of $ 20<f<100\,\mathrm{kHz} $), collapses at higher $ \bar{n}_e / n_G $ values, leading to a reduced Reynolds force $ \mathcal{F}_{Re} = - \partial_r \langle \tilde{v}_\theta \tilde{v}_r \rangle $.
The Reynolds power $ \mathcal{P}_{Re} = - \langle v_\theta \rangle \partial_r \langle \tilde{v}_\theta \tilde{v}_r \rangle $ (\cref{fig:vphase}(c)) can also be calculated, which is a measure of the nonlinear kinetic energy gained by the low-frequency sheared flow \cite{Tynan2013NF73053,Cziegler2014PPaCF75013,Yan2014PRL125002,Tynan2016PPaCF44003}.
Note here that this quantity looks at the net transfer of kinetic energy from the 20-100 kHz turbulent fluctuations into the low-frequency poloidal velocity ($ f<0.5 $ kHz).
The peak value of the Reynolds power decreases significantly, when $ \bar{n}_e /n_G $ is increased from \numrange{0.3}{0.8}, indicating a decline in the nonlinear kinetic energy transferred into the edge shear flow.

\begin{figure*}
\includegraphics{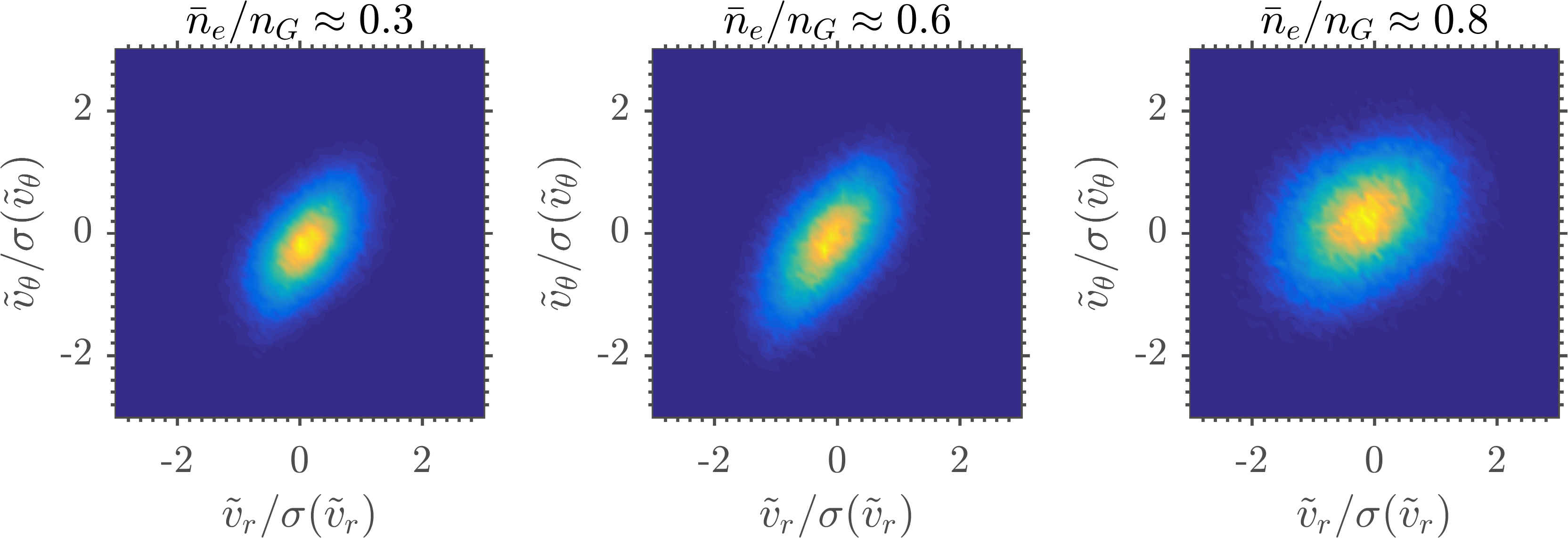}
\caption{\label{fig:pdf} Joint PDF of radial and azimuthal velocities, $\mathsf{P}(\tilde{v}_r, \tilde{v}_\theta)$, at $ r-r_{\rm sep} \approx -1 \, \mathrm{cm} $ at three densities. Velocities are normalized by their standard deviations.}
\end{figure*}

The turbulent drive for the sheared flow (Reynolds force $ \mathcal{F}_{Re} $) is, in principle, positively related to the eddy-tilting effect \cite{Manz2009PRL165004,Stroth2011PPaCF24006,Xu2011PRL55003}.
The eddy structures can be empirically represented by joint probability density functions (PDFs) of radial and azimuthal velocities \cite{Yan2008PoP92309}, i.e.~$ \mathsf{P}(\tilde{v}_r, \tilde{v}_\theta) \sim \langle k_{r} k_{\theta} \rangle $.
The contours of $ \mathsf{P}(\tilde{v}_r, \tilde{v}_\theta) $ at a position of $ r-r_{\rm sep} \approx -1 \, \mathrm{cm} $ at different normalized plasma densities are shown in \cref{fig:pdf}.
At lower densities, $ \mathsf{P}(\tilde{v}_r, \tilde{v}_\theta) $ is highly correlated and elongated along the diagonal direction.
As the density is raised to $ 0.8\,n_{G} $, $ \mathsf{P}(\tilde{v}_r, \tilde{v}_\theta) $ is more scattered and becomes more isotropic.
This observation is an indication of a reduced eddy-tilting effect by the sheared flow in high density plasmas.

\begin{figure}
\includegraphics[width=3.3in]{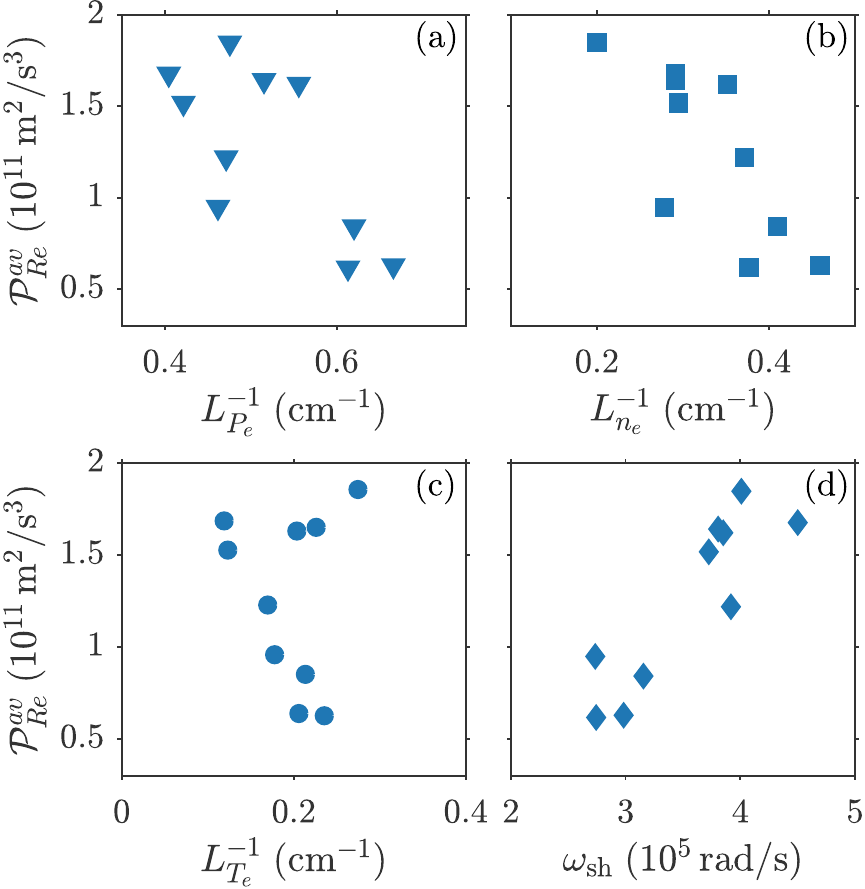}
\caption{\label{fig:av_Rey_pow} Averaged Reynolds powers, $ \mathcal{P}_{Re}^{av} = \int \mathcal{P}_{Re} \,rdr / \int rdr $ where $ -1< r - r_{\rm sep} < 1 $ \si{\centi\meter}, compare against the normalized pressure gradient $ L_{P_e}^{-1} $ (a), the normalized density gradient $ L_{n_e}^{-1} $ (b), the normalized electron temperature gradient $ L_{T_e}^{-1} $ (c), and the shearing rate of poloidal velocity (d).}
\end{figure}

Since the edge gradients provide free energy to the turbulence, it would be natural to seek the relation between relevant local gradients and the volume-averaged Reynolds power, $ \mathcal{P}_{Re}^{av} = \int \mathcal{P}_{Re} \,rdr / \int rdr $, where the integration is over $ -1 < r - r_{\rm sep} < 1 $ \si{\centi\meter}.
Figure \ref{fig:av_Rey_pow} shows the volume-averaged Reynolds power as a function of edge gradients: (a) normalized electron pressure gradient, $ L_{P_e}^{-1} = \partial_r \ln P_e $; (b) normalized density gradient $ L_{n_e}^{-1} = \partial_r \ln n_e $; (c) normalized electron temperature gradient, $ L_{T_e}^{-1} = \partial_r \ln T_e $; (d) mean shearing rate of poloidal velocity, $ \omega_\mathrm{sh} \approx \left| \frac{\partial \langle v_\theta \rangle}{\partial r} \right|$.
While there is no obvious linear dependence on the temperature gradient, $ \mathcal{P}_{Re}^{av} $ decreases as $ L_{n_e}^{-1} $ is increased, suggesting a suppression of the nonlinear energy transfer to the low-frequency shear flow in high density plasmas.

\begin{figure}
\includegraphics[width=3.3in]{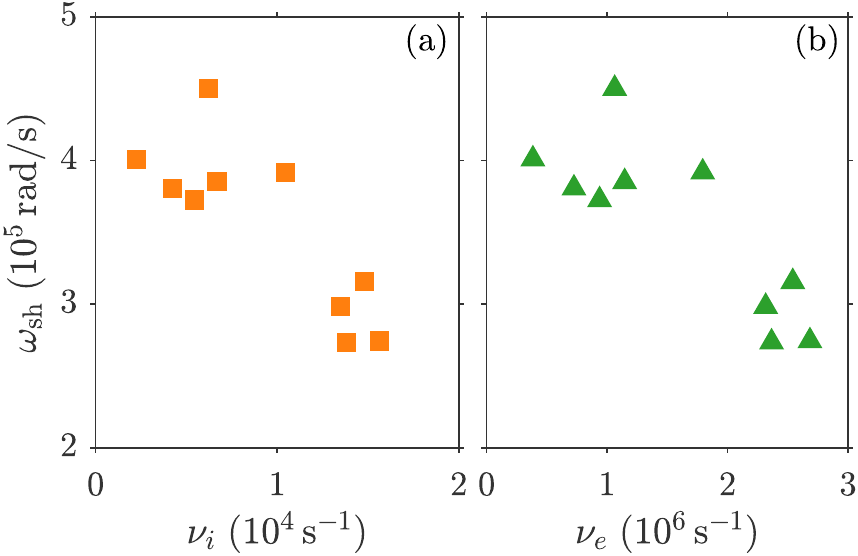}
\caption{\label{fig:shear_coll} The flow shearing rate, $ \omega_\mathrm{sh} $, compares against the ion collision rate $ \nu_\mathrm{i} $ (a) and the electron collision rate $ \nu_\mathrm{e} $ (b).}
\end{figure}

\begin{figure}
\includegraphics[width=3.3in]{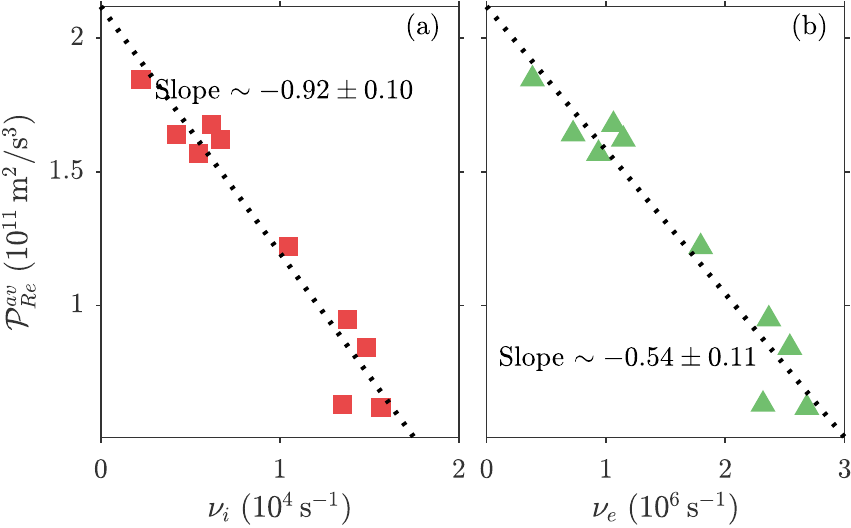}
\caption{\label{fig:Rey_coll} The averaged Reynolds power, $ \mathcal{P}_{Re}^{av} $, compares against the ion collision rate $ \nu_\mathrm{i} $ (a) and electron collision rate $ \nu_\mathrm{e} $ (b). Black dotted lines imply the linear trends.}
\end{figure}

The dependence on collision rates has also been studied. 
As shown in \cref{fig:shear_coll}, the shearing rate of the poloidal flow $ \omega_\mathrm{sh} $ decreases when the collision rate of either ions or electrons is raised.
Here, electron and ion collision rates are volume-averaged over $ -1 < r - r_{\rm sep} < \SI{1}{cm} $.
They are calculated respectively via $ \nu_\mathrm{e} = 2.91 \times 10^{-6} n T_e^{-3/2} \ln \Lambda $ and $  \nu_\mathrm{i} = 4.8 \times 10^{-8} Z^4 \mu^{-1/2} n T_i^{-3/2} \ln \Lambda $, with the approximation of $ T_i \approx T_e $, where $ Z $ is the charge number, $ \mu $ is the ion mass number, and $ \ln \Lambda $ is the Coulomb logarithm, which is 13.6 for electrons and 6.8 for ions.
This phenomenon conforms to the prediction \cite{Lin1999PRL3645,Diamond2005PPaCF35} that stronger Coulomb collisions damp zonal flows at higher collisionality.
Correspondingly, the averaged Reynolds power $ \mathcal{P}_{Re}^{av} $ decreases with increasing collision rates (\cref{fig:Rey_coll}), indicating that the nonlinear energy transfer to the edge shear flow is reduced at higher collision rates.

\begin{figure}
  \includegraphics[width=3in]{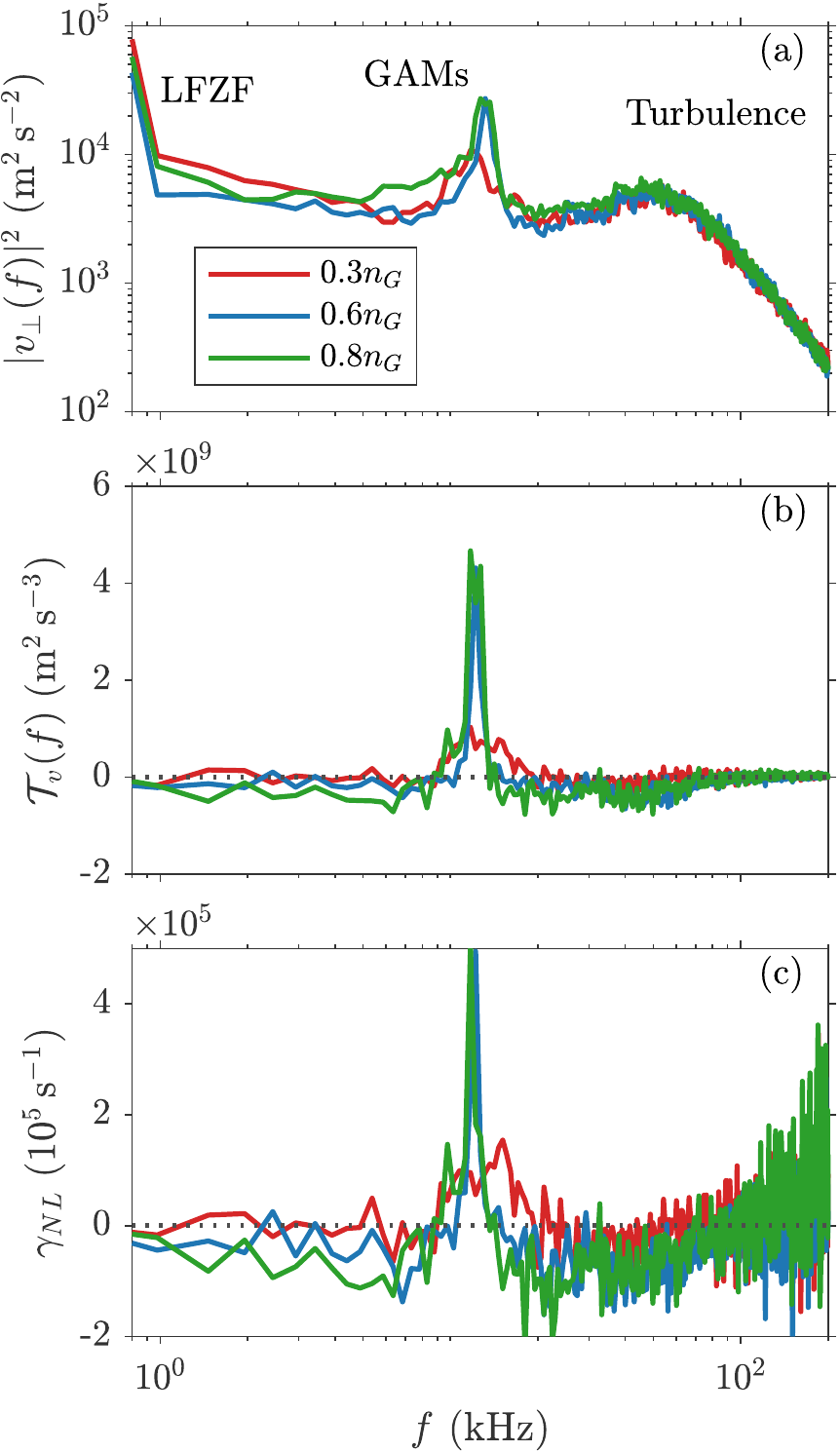}
  \caption{\label{fig:bispec1} Auto-spectra of perpendicular velocity fluctuations (a), the nonlinear kinetic energy transfer rate $ \mathcal{T}_v(f) $ (b), and the effective growth rate due to energy transfer $ \gamma_\mathrm{NL} = \mathcal{T}_{v}(f)/\langle \tilde{\mathbf{v}}_\perp^2(f) \rangle $ (c), at three normalized densities $ \bar{n}_e / n_G $, at a position of $ r-r_{\rm sep} \approx -2 \, \mathrm{cm} $.}
\end{figure}

The kinetic energy transfer between the edge turbulence and shear flows has also been investigated in the frequency domain.
In the auto-spectra of perpendicular velocities $ \mathbf{v}_\perp (f) $ (\cref{fig:bispec1}(a)), at least two distinct flow patterns can be recognized, which are geodesic acoustic modes (GAMs) (centered at $ f \approx 12 \,\mathrm{kHz} $) and the turbulence ($ f > 30 \,\mathrm{kHz} $). These two patterns have been observed in previous experiments in this device \cite{Lan2008PPaCF45002,Xu2012PRL245001,Zhao2006PRL255004}.
While there is no obvious changes in the spectra of turbulent velocities, the power contained in GAMs velocity fluctuations increases by a factor of three as $ \bar{n}_e /n_G $ is raised from 0.3 to 0.8.

\begin{figure}
\includegraphics[]{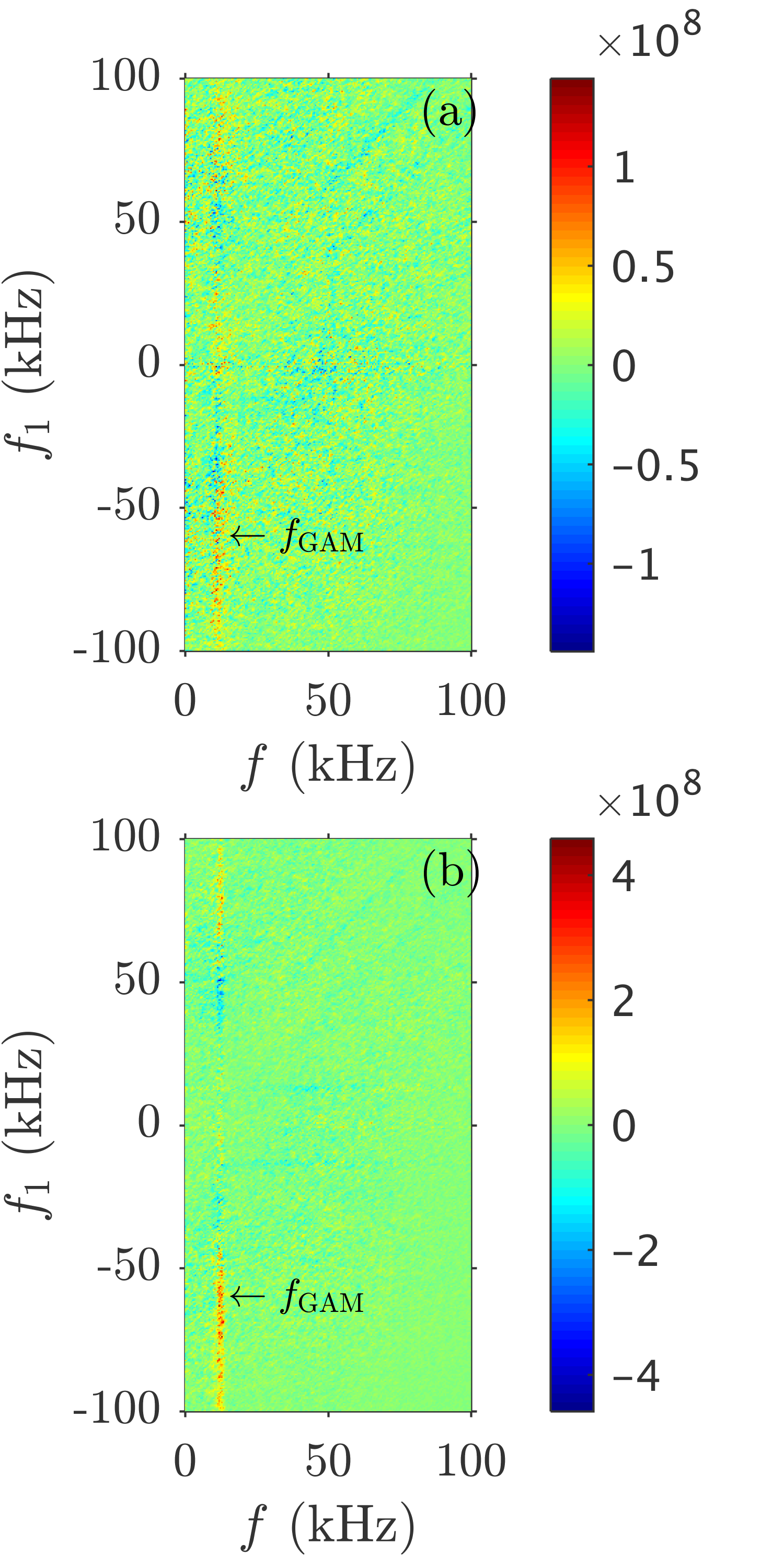}
\caption{\label{fig:bispec2} 2D nonlinear kinetic energy transfer for $ \bar{n}_e / n_G \approx 0.3 $ (a) and $ 0.8 $ (b). A positive value (red) at $ (f, f_1) $ means that the perpendicular velocity fluctuations, $ \mathbf{v}_\perp $, associated with $ f $ gain kinetic energy from those at $f_1$; a negative value (blue) means $ \mathbf{v}_\perp $ at $ f $ lose energy to those at $f_1$. Clearly, the fluctuations at $ f_{\rm GAM} \approx 12 $ kHz gain energy from ambient turbulence ($ 40<f<100 $ kHz).}
\end{figure}

The 2D frequency-resolved nonlinear energy transfer, $ \mathcal{T}_{v}(f, f_1) = \langle \mathbf{v}^{*}_{f} \cdot (\mathbf{v}_{f-f_1} \cdot \nabla_\perp \mathbf{v}_{f_1}) \rangle $, for $ \bar{n}_e / n_G = 0.3 $ and $ 0.8 $ are shown in \cref{fig:bispec2}, which are computed from 100 ensembles of time-stationary data taken roughly at $ r-r_{\rm sep} = -2\, \mathrm{cm}$.
A positive value (red) at $ (f, f_1) $ suggests that the perpendicular velocity fluctuations associated with $ f $ gain kinetic energy from fluctuations at $ f_1 $; a negative value (blue) suggests that the fluctuations at $ f $ lose energy to those at $ f_1 $.
More detailed description of this method can be found in earlier publications \cite{Xu2010PoP32311,Xu2012PRL245001}.
As shown in \cref{fig:bispec2}, the GAMs (at $ f\approx 12 \,\mathrm{kHz}$) gain energy from high-frequency fluctuations ($ f\approx 40-100\, \mathrm{kHz}$).
Figure \ref{fig:bispec1}(b) shows the the net frequency-resolved nonlinear energy transfer rate \cite{Xu2012PRL245001,Xu2010PoP32311}, $ \mathcal{T}_v(f) = - \Re \sum_{f_1} \left< \mathbf{v}_{\perp,f}^* \cdot \left( \mathbf{v}_{\perp,f-f_1} \cdot \nabla_\perp \mathbf{v}_{\perp,f_1} \right) \right> $, at different $ \bar{n}_e /n_G $ values, which can be obtain by integrating over $ f_1 $ axis in the 2D nonlinear energy transfer map.
The GAMs appear to gain more kinetic energy from the turbulent fluctuations when $ \bar{n}_e / n_G $ is higher.
By normalizing the energy transfer rate using auto-power of perpendicular velocity fluctuations, we can obtain the effective frequency-resolved nonlinear growth or damping rate (\cref{fig:bispec1}(c)), $ \gamma_\mathrm{NL}(f) = \mathcal{T}_v(f)/\langle \tilde{\mathbf{v}}_\perp^2(f) \rangle $.
As shown in \cref{fig:bispec1}(c), the effective nonlinear growth rate of GAMs, $ \gamma_\mathrm{NL}^\mathrm{GAM} $, increased significantly as $ \bar{n}_e / n_G $ is raised.

\begin{figure}
\includegraphics[width=3in]{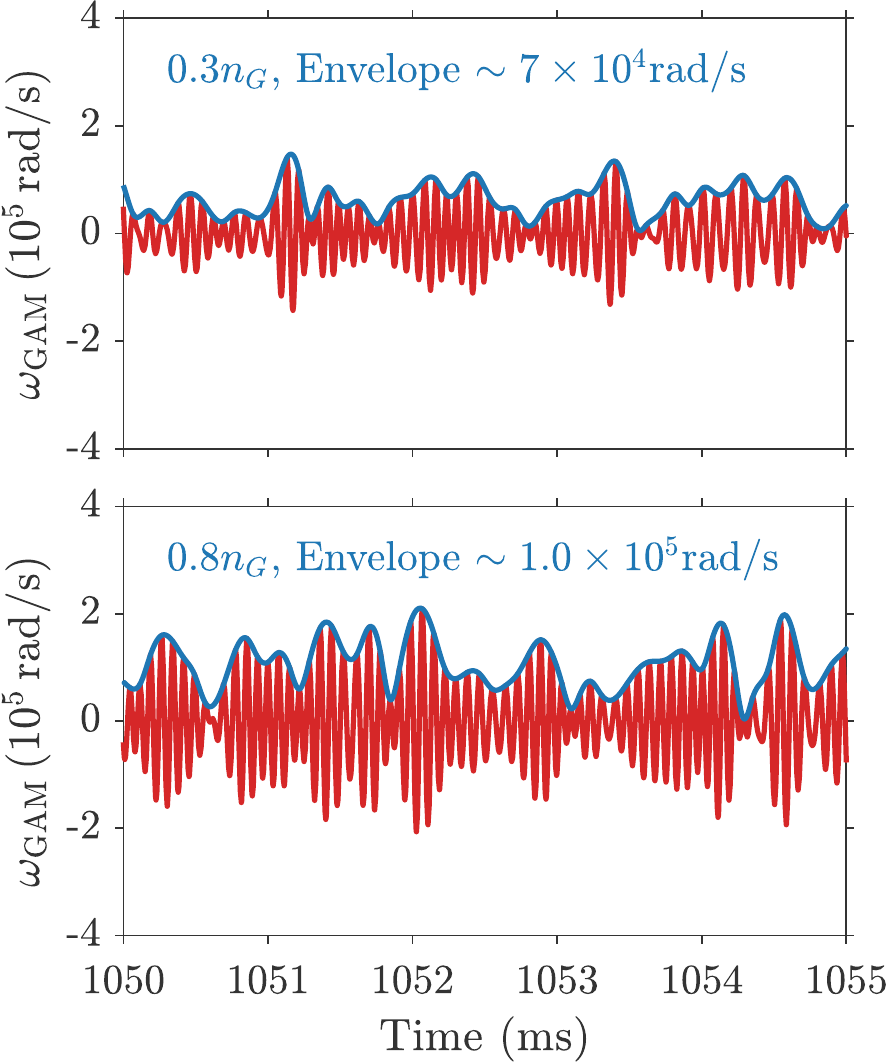}
\caption{\label{fig:GAM_shearing} The shearing rate of GAMs, $ \omega_{\rm GAM} $, for $ \bar{n}_e / n_G \approx 0.3$ (upper) and 0.8 (bottom). Blue curves indicate envelopes of $ \omega_{\rm GAM} $. The mean values of envelops are $ \sim 7\times10^4 \,{\rm rad/s}$ for $ \bar{n}_e / n_G \approx 0.3$ and $ \sim 10\times10^4 \,{\rm rad/s}$ for $ \bar{n}_e / n_G \approx 0.8$.}
\end{figure}

The shearing rate of GAMs can be estimated via $ \omega_{\rm GAM} = \partial_r v_{\theta}^{\rm GAM}$, where the GAM velocity $ v_{\theta}^{\rm GAM} $ is filtered into the frequency range of $ 9<f<15 $ kHz using a fifth-order Butterworth filter. 
As shown in \cref{fig:GAM_shearing}, the mean value of $ \omega_{\rm GAM} $'s envelope increases from 7 to \SI{10e4}{\per\second} when $ \bar{n}_e / n_G $ increases from 0.3 to 0.8.
Also, the eddy turn-over rate is estimated as $ \omega_{\rm eddy} = \tau^{-1}_{\rm eddy} \sim \frac{\tilde{\phi}_f}{B L_r L_\theta} \sim 4.6 - \SI{12e4}{\per\second}$, where $ B = \SI{1.3}{\tesla} $ is the toroidal field, and $ \tilde{\phi}_f \sim 30-\SI{50}{\volt}$ is the fluctuation amplitude of  floating potentials, and $ L_r \sim \SI{1}{cm} $ and $ L_\theta \sim 3-\SI{5}{cm}$ are respectively the correlation lengths in radial and poloidal directions \cite{Zhao2006PRL255004,Xu2012PRL245001}.
While the shearing rate of GAMs is comparable to the eddy turn-over rate at higher densities, i.e.~$ \omega_{\rm GAM} \sim \omega_{\rm eddy} $, it is still less than the mean flow shearing rate, i.e.~$ \omega_{\rm GAM} \sim 0.3 \times \omega_{\rm sh}$.
These findings suggest that mean flow plays the leading role in turbulence suppression.

\subsection{Enhanced Particle Transport}

\begin{figure}
\includegraphics[width=3in]{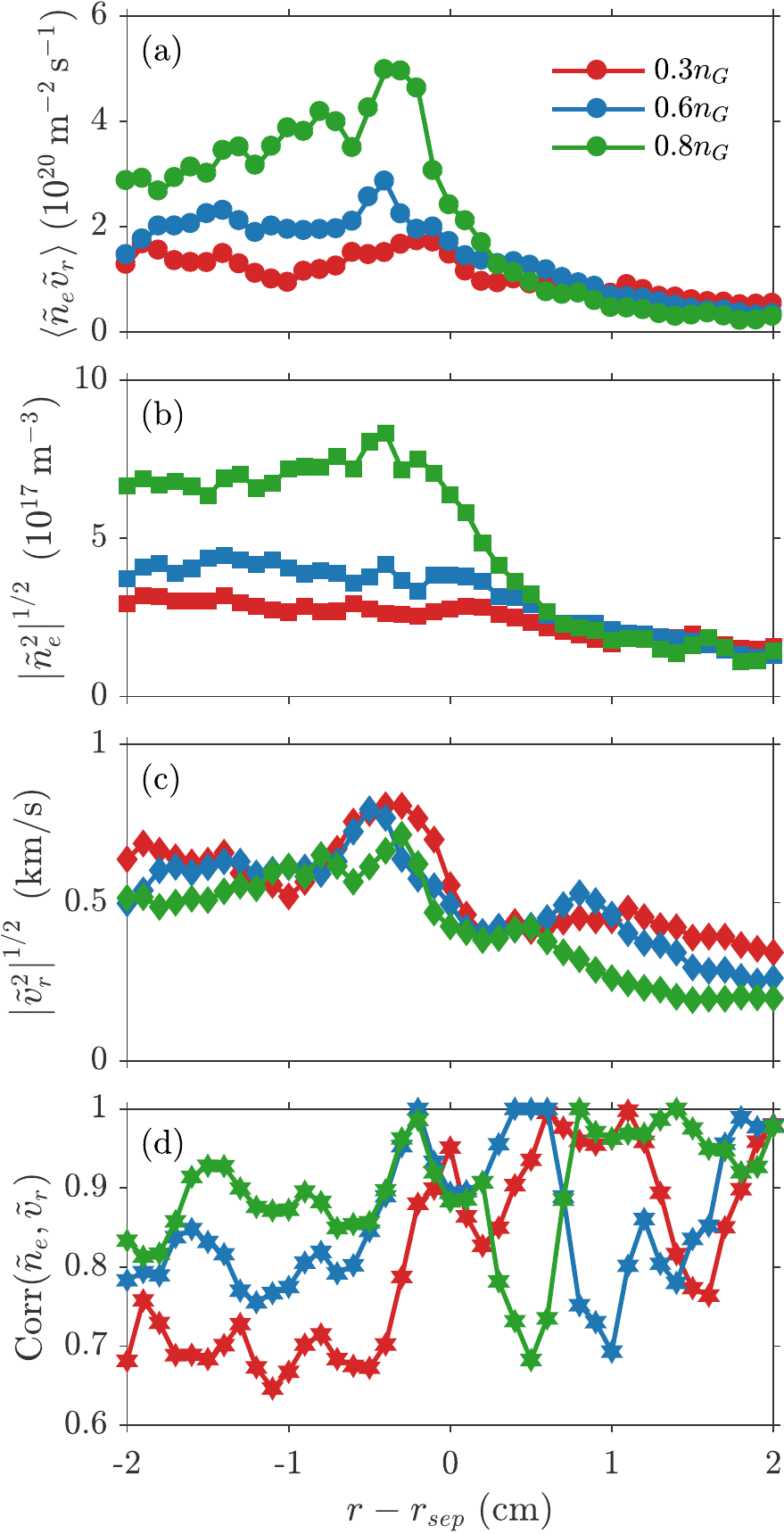}
\caption{\label{fig:part_flux} Radial profiles of electron particle flux (a), RMS of electron density fluctuations (b), RMS of radial velocity fluctuations (c), and cross-correlation between velocity and density fluctuations (d).}
\end{figure}

Figure \ref{fig:part_flux}(a) shows the radial profiles of particle flux at three normalized core densities.
The turbulent particle flux, $\Gamma_r = \langle \tilde{n}_e  \tilde{v}_r \rangle$, increases substantially when $ \bar{n}_e / n_G $ is raised from 0.3 to 0.8, in spite of the increase of GAMs amplitudes.
The root-mean-square (RMS) of the density and radial velocity fluctuations ($ 20<f<100\,\mathrm{kHz} $) are shown in \cref{fig:part_flux}(b) and \ref{fig:part_flux}(c), respectively.
While the variation in RMS of radial velocity fluctuations is negligible, the RMS of electron density fluctuations grows by a factor of two as the core density is increased.
The cross correlation coefficient $ {\rm Corr}(\tilde{n}_e,\tilde{v}_r) $  also increases with $ \bar{n}_e / n_G $ values inside the separatrix (\cref{fig:part_flux}(d)).
Here, the cross correlation coefficient at each position between $ \tilde{n}_e $ and $ \tilde{v}_r $ is evaluated via $ {\rm Corr}(\tilde{n}_e,\tilde{v}_r) = \frac{\langle \tilde{n}_e \tilde{v}_r \rangle}{\sigma_{n_e}\sigma_{v_r}} $, where $ \sigma_{n_e} $ and $ \sigma_{v_r} $ are standard deviations of density and radial velocity fluctuations, respectively.

\begin{figure}
  \includegraphics[width=3in]{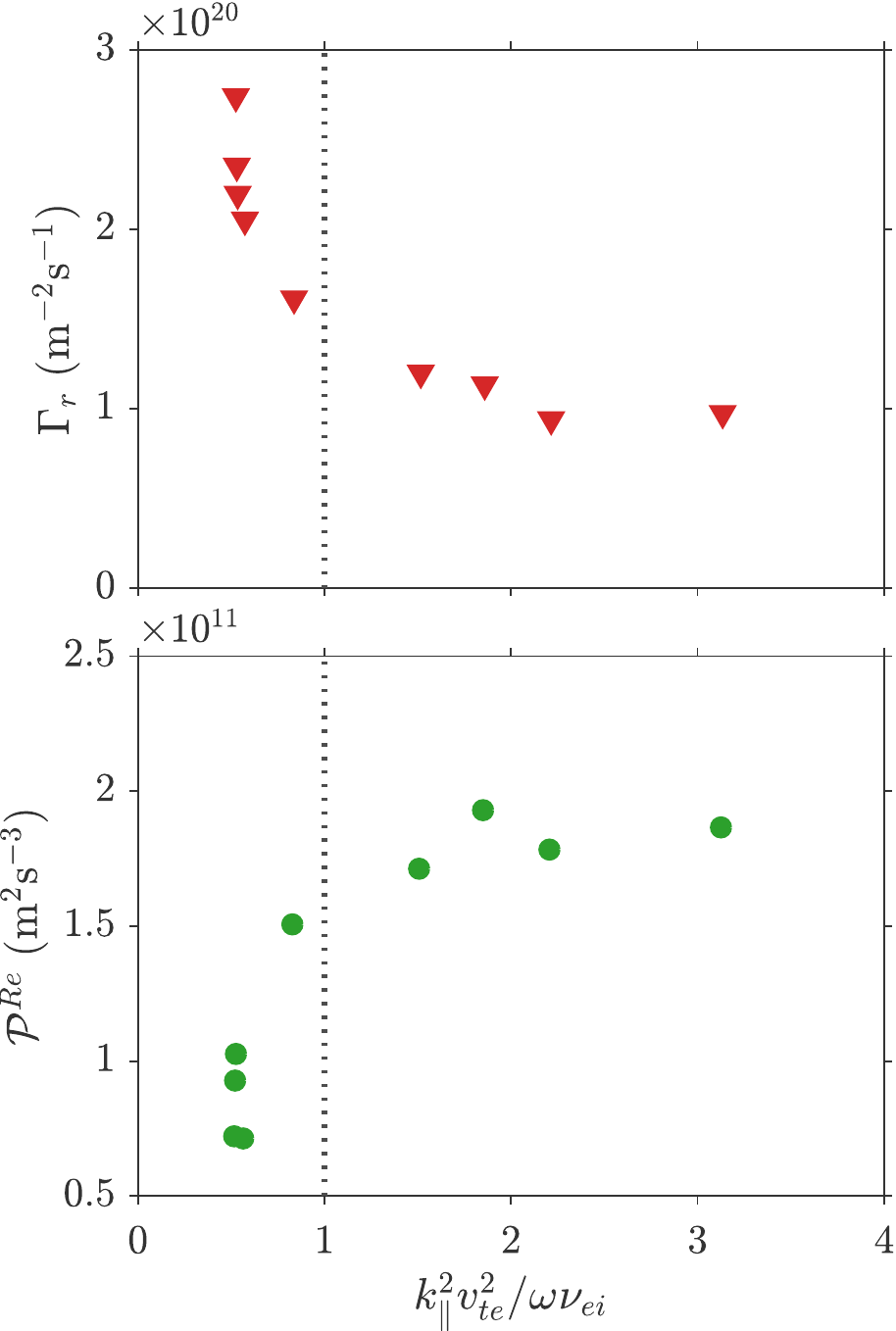}
  \caption{\label{fig:adia} The volume averaged particle flux (upper) and Reynolds power (bottom) plotted as a function of the adiabatic parameter.}
\end{figure}

During the density scan, the adiabatic parameter, $ k_\parallel^2 v_{te}^{2} / \omega\nu_e $, is estimated to drop from about 3 to 0.5, where $ k_\parallel \sim 1/qR $ is the parallel wavenumber, $ v_{te} $ is the electron thermal speed, $ \nu_e $ is the electron collision rate, and $ \omega $ is the dominant frequency of turbulence.
Such substantial change in $ k_\parallel^2 v_{te}^{2} / \omega\nu_e $ can induce a non-adiabatic electron response \cite{Stroth2011PPaCF24006,Scott2005NJoP92}, i.e.~the dominant modes may switch from adiabatic drift waves ($ k_\parallel^2 v_{te}^{2} / \omega\nu_e > 1 $) to non-adiabatic resistive driven modes ($ k_\parallel^2 v_{te}^{2} / \omega\nu_e < 1 $).
As shown in \cref{fig:adia}, with decreasing adiabaticity, the edge particle transport $ \Gamma_r $ rises by a factor of three.
Here, the edge particle transport is represented by the volume-averaged particle flux, $ \langle \Gamma_r \rangle = \int \langle \tilde{n}_e  \tilde{v}_{r} \rangle \,rdr / \int rdr$ inside the separatrix ($-2<r-r_{\rm sep}<0\,\mathrm{cm} $).
Concurrently, the volume-averaged Reynolds power drops significantly when adiabaticity is less than one.

\section{Discussion}
\label{sec:disc}

\subsection{Reduced Shear Flows and Enhanced Particle Transport}

\begin{figure}
  \includegraphics[width=3in]{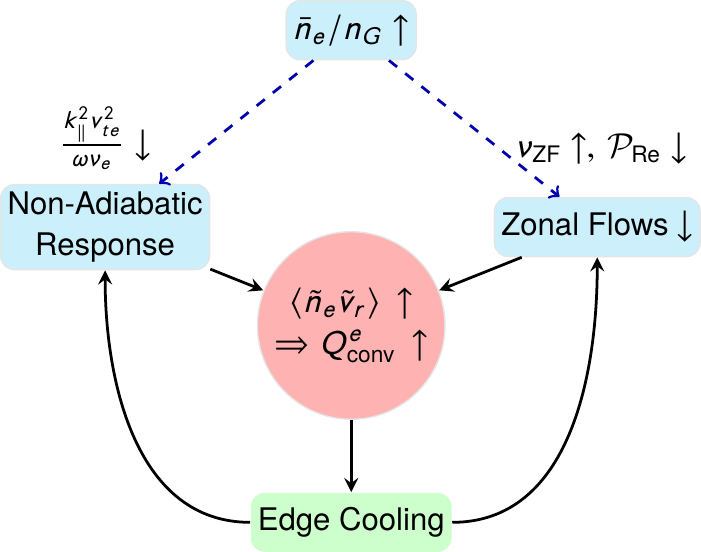}
  \caption{\label{fig:coll_flow} Sketch of a possible feedback mechanism in high density plasmas. The increased collision rate increases the collisional dissipation of zonal flows, and enhances the correlation between $ \tilde{n}_e $ and $ \tilde{v}_r $. Both effects can further enhance turbulent particle flux and edge cooling.}
\end{figure}

One of the main goals of this study is to understand the evolution of edge sheared flows and their effects on regulating particle transport as the Greenwald density limit is approached.
As the line-averaged density is raised, the collisional dissipation of the low-frequency zonal flow $ \nu_{\rm ZF} $ increases and the Reynolds power $ \mathcal{P}^{Re} $ collapses (\cref{fig:coll_flow}).
As a result, low-frequency zonal flows are strongly damped and can no longer mitigate turbulent particle transport.
The enhanced particle losses result in a drop in edge electron temperature which in turn further reduced the zonal flow and its turbulent drive.
This process iterates via a closed feedback loop and leads to the development of edge cooling.
This picture is opposite to the \textit{L-H} transition physics \cite{Diamond2005PPaCF35,Fujisawa2009NF13001,Stroth2011PPaCF24006,Tynan2016PPaCF44003} in which the turbulent transport is suppressed by zonal flows that in turn is driven by the turbulence via the Reynolds force.
Here, the eddy-tilting and therefore the Reynolds force are reduced as collision rates are increased.

In addition to effects of shear flows and Reynolds force, particle transport can also be altered by the non-adiabatic electron response (\cref{fig:coll_flow}).
The significant drop in the adiabatic parameter $ k_\parallel^2 v_{te}^{2} / \omega\nu_e $ suggests a conversion from \emph{adiabatic} drift waves to \emph{non-adiabatic} resistivity driven modes, e.g.~resistive ballooning modes, due to the increased collision rate \cite{Thayer1987PoF3724}.
Theoretical models predict that edge turbulent transport can be enhanced by resistive ballooning modes when the plasma density exceeds the critical level \cite{Rogers1998PRL4396,Xu2003PoP1773,Scott2005NJoP92,Tokar2003PRL95001,Tokar2005PoP52510}.
The increased particle losses and heat flux may also trigger the cooling of edge plasmas, as shown in \cref{fig:coll_flow}.
The edge cooling then increases the current density gradient and drives the MHD instabilities to which the radiative models are applicable.
The measurements from SOL region of Alcator C-Mod \cite{LaBombard2005NF1658} show that cross-field transport increases dramatically with increasing collisionality, and is qualitatively in agreement with a ``density-limit boundary'' predicted by the theoretical models \cite{Rogers1998PRL4396,Xu2003PoP1773}.
Moreover, in H-mode plasmas the Reynolds stress is too weak to drive the zonal flows.
Thus, the resistive modes induced transport and increased collisional dissipation of sheared flows are more relevant mechanisms for the H-mode density limit.

In order to obtain steady edge profiles, we employed the shot-by-shot density scans in this study.
However, a perturbative study using the density ramp-up or modulation would be necessary to resolve which is first affected by the increased collision rates, the collisional damping of shear flows or the collapse of the Reynolds force (turbulent vorticity flux).

Apart from the poloidal shear flows discussed in the present study, the toroidal shear flows can play an important role in mitigating instabilities and improving the plasma confinement.
Accordingly, the dynamics of toroidal shear flows and their coupling with poloidal flows in high collisionality plasmas may deserve further explorations.

\subsection{Different Behaviors of Zonal Flows and GAMs}

The results shown in \cref{subsec:eng} demonstrate that the low-frequency zonal flow gains less energy from the turbulence at higher densities, and its shearing rate decreases as well.
On the other hand, GAMs gain more energy and thus have higher effective growth rate $ \gamma_\mathrm{NL}^\mathrm{GAM} $ at higher densities, even though the turbulence intensity does not change.
Similar observations on GAMs have been reported in a recent investigation from JET \cite{Silva2016NF106026}, in which GAMs amplitudes measured by Doppler backscattering increase as the line-averaged density is raised.
The competition between ZFs and GAMs has also been observed in earlier experiments in HL-2A \cite{Xu2012PRL245001} and Alcator C-Mod \cite{Cziegler2015NF83007,Cziegler2017PRL105003}.
In HL-2A's ECRH power scanning experiments, the amplitudes and effective nonlinear growth rates $ \gamma_\mathrm{NL} $ of ZFs and GAMs were found to increase with the ECRH power, until the ZFs dominate the nonlinear energy transfer process when $ P_\mathrm{ECRH} \geq 730\,\mathrm{kW} $ \cite{Xu2012PRL245001}.
Although some theoretical models \cite{Miki2010PoP32309,Miki2011NF103003} have explored the different behaviors of low-frequency ZFs and GAMs, a detailed comparison between the theory and measurements is still lacking.
The physics of the coupling between ZFs and GAMs as a function of the heating power and plasma density remains to be studied.

\subsection{Potential Effect of Magnetic Stress}

One topic that deserves further investigation is the effect of magnetic stress, $  \langle \tilde{B}_{\theta} \tilde{B}_{r} \rangle $, on the driving force for zonal flows near the Greenwald limit.
The divergence of the Maxwell stress is known to induce a force on plasmas.
The signs of the divergences of the Reynolds stress and magnetic stress are opposite for the drift-Alfven waves \cite{Diamond2005PPaCF35,Kleva2008PoP82307}, resulting in a lower driving force for the zonal flows in the limit of finite $ \hat{\beta} $.
As reported in both experiments \cite{LaBombard2005NF1658} and numerical simulations \cite{Rogers1998PRL4396}, electromagnetic fluid drift turbulence grows and becomes the dominant modes controlling edge transport when the density limit is approached.

In present study, the MHD ballooning parameter, $ \alpha_{\rm MHD} = \frac{q^2 R}{L_{Pe}} \beta $ with $ \beta = \frac{4 \mu_{0} P_{e0}}{B^2} $, increases from about 0.1 to 0.3 at the edge as $ \bar{n}_e / n_G $ is raised from 0.3 to 0.8.
Therefore, magnetic fluctuations are supposed to increase, and their effects on shear flows should be considered.
Nonetheless, even without any direct measurement of electromagnetic effects, the reduction in turbulent force for the zonal flows at higher densities suggests that zonal flow is an important element in density limit physics.
A probe array that is capable of measuring magnetic and Reynolds stresses has been developed.
Direct magnetic stress measurements are in progress.
We hope to report more results on this topic in the future.

\section{Conclusion}
\label{sec:concl}

Using a multi-tip Langmuir probe array, edge turbulent particle transport and shear flows have been investigated as the Greenwald limit is approached in the HL-2A tokamak.
As the line-averaged density increases toward the Greenwald limit, the low-frequency zonal flow (ZF) shear and its turbulent drive (Reynolds power) are observed to decrease with increasing collision rates.
The eddy-tilting and Reynolds force are reduced, thus ZF cannot regulate turbulent transport efficiently.
The GAMs gains more energy from the ambient turbulence at higher $ \bar{n}_e/n_G $ values, but do not mitigate the turbulent particle transport.
On the other hand, the adiabatic parameter, $ k_\parallel^2 v_{te}^{2} / \omega\nu_e $, drops significantly from about 3 to 0.5 as $ \bar{n}_e / n_G $ increases from 0.3 to 0.8.
This substantial decrease in adiabaticity is associated with both reduced Reynolds power and enhanced edge particle flux.
These findings suggest that as the Greenwald density limit is approached, the increased collision rates may not only induce non-adiabatic electron response, but is also associated with a decrease in the low-frequency zonal flow and its turbulent drive.
Both effects can give rise to enhanced edge particle transport and thus edge cooling.

\section*{Acknowledgments}
This work is supported by the Chinese National Fusion Project for ITER under Grant No 2013GB107001, the National Natural Science Foundation of China under Grant Nos 11375053 and 11575055, and the International S\&T Cooperation Program of China under Grant No 2015DFA61760.

\bibliography{density_limit}

\end{document}